\documentclass{nature}
\usepackage{subfig}
\usepackage[colorlinks=true,allcolors=blue]{hyperref}
\usepackage[colorinlistoftodos]{todonotes}
\usepackage{float}
\usepackage{lineno}
\usepackage{color}
\usepackage{graphicx,dblfloatfix}
\usepackage[labelfont=bf]{caption}
\DeclareCaptionLabelSeparator{bar}{\textbar}

\DeclareCaptionFont{compact}{%
  \footnotesize
  \selectfont
}

\usepackage{amsmath}
\usepackage{amssymb}
\usepackage{chapterbib}
\newcommand{\blue}{\textcolor{blue}}

\title{Mie-lithography: void-resonance-assisted femtosecond laser printing for deep-ultraviolet to near-infrared nano dispersion devices}

\author{Wei Gong$^{1,\dagger}$, Zhen-Ze Li$^{2,\dagger,*}$, Chang Yu$^1$, Zhen Wang$^{1}$, Han-Yue Fan$^{1}$, Yi Wang$^{2}$, Zhi-Hao Chen$^{1}$, Chun-Qi Jin$^{1}$, Yu-Hao Lei$^{3}$, Qi-Dai Chen$^{1,*}$, Lei Wang$^{1,*}$, and Hong-Bo Sun$^{1,2,*}$}

\begin{document}

\maketitle

\begin{affiliations}
 \item State Key Laboratory of Integrated Optoelectronics, College of Electronic Science and Engineering, Jilin University, Changchun 130012, China
 \item State Key Laboratory of Precision Measurement Technology and Instruments, Department of Precision Instrument,
 Tsinghua University, Beijing
100084, China
 \item MIIT Key Laboratory of Complex-field Intelligent Exploration, School of Optics and Photonics, Beijing Institute of Technology, Beijing 100081, China

 \item[$^\dagger$] These authors contributed equally to this work

 \item[$^*$] Corresponding authors:\\
  Zhen-Ze Li: zzli2023@mail.tsinghua.edu.cn\\
  Qi-Dai Chen:
  chenqd@jlu.edu.cn\\
  Lei Wang:
  leiwang1987@jlu.edu.cn\\
  Hong-Bo Sun: hbsun@tsinghua.edu.cn

\end{affiliations}

\clearpage

\begin{abstract}
Nanoscale control of optical dispersion is essential for applications ranging from miniaturized spectrometers to color printing, all of which demand broadband spectral tunability. However, the Kramers–Kronig relations impose a fundamental trade-off between dispersion and loss, strictly limiting the design ability of single-material devices across the deep-ultraviolet (DUV) to near-infrared (NIR) regimes. Consequently, the fabrication of miniaturized dispersion devices heavily relies on costly nanofabrication or heterogeneous integration. Here we overcome these limitations by shifting the light–matter interaction from solid structures into air-filled voids. We introduce a fabrication strategy termed ‘Mie-lithography’, in which laser-printed seed nanocavities excite Mie resonances in air and the resulting localized field enhancement drives the self-assembly of three-dimensionally tunable void-type optical resonators. Because the resonant modes are primarily confined within air voids, this architecture effectively alleviates material-imposed dispersion–loss constraints, allowing on-demand customization of the broadband spectral response. This approach could enable single-step, high-throughput ($\ge 10^6$~pixels/s) printing of dispersion units with a resolution of 63,500 DPI. As a proof of concept, we demonstrate a single-material nano dispersion device for spectral encoding from 200 to 800~nm, with spectral reconstruction using a camera. Our approach can be extended into a platform for ultra-broadband (DUV-NIR) nanodevices fabrication and design, opening avenues for high-pixel-density displays and miniaturized spectrometers.
\end{abstract}

\newpage


\noindent Optical dispersion at the nanoscale represents a central pursuit in nanophotonics\cite{wang2023structural,yang2019structural,li2024recent}. Since Lord Rayleigh’s discovery that the color of the sky originates from atmospheric nanoparticle scattering, it has been recognized that nanostructures supporting optical resonances (such as Mie resonance) can enhance near-field effects controlling far-field scattering through
 intermodal interference. This capability has, in turn, fueled the development of miniaturized dispersion devices, with applications spanning sensing\cite{shi2022multifunctional,arslan2025attoliter,lin2017large}, subwavelength lasers\cite{zeng2025metalasers,liu2019high,luan2025all,di2024all,liu2018all}, structural colors\cite{yang2020all,hentschel2023dielectric,jang2020spectral,hu2018laser}, and data storage\cite{wiecha2019pushing,li2022metasurface}. However, translating these resonant structures into practical, broadband devices, e.g., from deep-ultraviolet (DUV) to near-infrared (NIR), faces what has long been considered an insurmountable challenge\cite{liu2020dielectric,yang2024advanced,chen2020flat}, rooted in both materials science and fabrication technology.

The challenge is twofold. The first is a fundamental physical limit imposed by the well-known Kramers-Kronig relations\cite{lucarini2003dispersion,landau2013electrodynamics}. Since Mie resonance typically occurs when the wavelength of light inside the nanoparticle ($\lambda_0/n$) approaches its spatial dimensions\cite{kivshar2022rise,kuznetsov2016optically,wriedt2012mie}, a high refractive index $n$ is preferred to enhance the resonant subwavelength light localization\cite{kuznetsov2016optically}. However, the extinction coefficient of an optical material is determined by the integration of the refractive index, $k\left(\omega\right)\propto\int{\frac{n\left(\omega^\prime\right)-1}{\omega^{\prime2}-\omega^2}\mathrm{d}\omega^\prime}$, and no single material can escape the intrinsic trade-off between strong light-matter interaction and low loss across such a vast spectral range. Regions of desirable high refractive index or strong dispersion are invariably linked to prohibitive absorption peaks. Consequently, materials such as silicon, which are commonly used in near-infrared and visible devices, are not viable in the ultraviolet range due to their strong absorption, and must be replaced with alternative materials such as hafnium dioxide (HfO$_2$)\cite{zhang2020low} or zirconium dioxide (ZrO$_2$)\cite{kim2023one}. A promising route toward broadband devices has been demonstrated using inverse dielectric resonators, commonly referred to as Mie voids. By shifting the resonance from the solid material into an air-filled cavity, Mie voids reduce the modal overlap with the lossy and dispersive host material. Previous studies have demonstrated Mie voids not only in the visible spectral range but also in the deep-ultraviolet region, where multiple higher-order resonant modes were observed\cite{hentschel2023dielectric}. Building on this demonstration, subsequent studies have extended Mie voids to sensing\cite{niinomi2025mie,arslan2025attoliter}, structural color\cite{hentschel2023dielectric,arslan2026polarization}, and light emission\cite{liao2026mie}.
The second constraint arises from the fabrication challenges of ultra-broadband nano dispersion devices. DUV resonators often require feature sizes approaching or below $100~$nm; for example, HfO$_2$-based DUV nanostructures typically involve electron-beam lithography, atomic layer deposition, ion milling, and resist removal\cite{zhang2020low}. Integrating different material systems across spectral bands adds deposition, patterning, and alignment steps, thereby increasing fabrication costs and imposing process-compatibility constraints related to deposition temperatures and material interfaces; a recent broadband hyperspectral imager, for instance, combined 16 organic materials through multiple deposition and photolithography cycles\cite{bian2024broadband}. Moreover, mask-based lithography efficiently replicates fixed designs but offers limited design flexibility, whereas electron-beam lithography and focused-ion-beam milling enable arbitrary patterning at low throughput\cite{leng2024meta,yang2024advanced}. These requirements complicate the scalable and low-cost fabrication of DUV–NIR dispersion devices.

In this article, inspired by the above-mentioned advances in Mie voids, we show that the solution to these dual challenges is rooted in a single, unifying principle: shifting the resonant light-matter interaction from the solid material into an air-filled void, from which a new fabrication method and a new device architecture can be developed simultaneously. First, we introduce a manufacturing strategy termed Mie-lithography (Mie void laser drilling), which harnesses the Mie resonance itself as an active nanofabrication tool. Its physical principle is a laser-ablation feedback process enabled by ``resonance-in-air'': (1) an initial, laser-ablated nanocavity acts as a seed structure; (2) subsequent laser pulses excite strong Mie resonances within this seed, creating an intensely localized field within the cavity that selectively drives further material ablation; (3) The evolving geometry of the cavity continuously reshapes the resonant mode, which in turn guides the ablation until the resonant wavelength of the cavity aligns with that of the laser. This dynamic interplay between the light field and the structure enables the controlled, self-assembled growth of three-dimensional void-type resonators (including transverse geometries and longitudinal depths). On the other hand, by creating these void-type resonators, the resonant medium is actually shifted from a lossy, dispersive solid (such as silicon) to air, which effectively alleviates the dispersion-loss trade-off, and allows the dimensions of resonators to be scaled to the wavelength level, further significantly relaxing the fabrication demands.

As a proof of concept, we demonstrate Mie-lithography-based color printing on high refractive-index materials with a wide gamut, achieving a resolution of up to 63,500 DPI with a projected printing speed $\ge 10^6$ pixels/s. Moreover, by leveraging the dispersion behavior of Mie resonators of varying sizes, we fabricated a compact spectral chip spanning 200–800~nm---to the best of our knowledge, exceeding the bandwidth achievable by any current miniaturized device (Fig.~\ref{F4}). Owing to the single-step nature of the fabrication process, our approach is readily scalable to other ultra-compact broadband dispersion devices and offers a viable route toward high-throughput and low-cost manufacturing.

\subsection{Self-assembly of Mie resonators via Mie-lithography} In this section, we elucidate the fabrication principle of Mie-lithography. By combining theoretical analysis and experimental evidence, we reveal that the key mechanism is a feedback loop between the incident laser beam and the evolving nanocavity, a process uniquely enabled by this ‘resonance-in-air’ concept (Fig.~\ref{F1}a).

The process begins when a tightly focused laser pulse (see Methods) creates an initial subwavelength cavity via optical breakdown. This cavity, a low-refractive-index region ($n_{air}\approx1$) embedded in a high-index substrate (e.g., silicon), acts as the dual of a solid nanoparticle according to the quasi-Babinet principle\cite{hamidi2025quasi}. Consequently, its scattering behavior is also governed by the ratio of its size ($d$) to the wavelength ($\lambda$) of the incident laser beam. When $d$ is significantly smaller than the wavelength ($d \ll \lambda$), the cavity exhibits a Rayleigh scattering behavior and the optical near-field distribution near the scatterer manifests as a dumbbell-shaped profile perpendicular/parallel to the laser polarization (Fig.~\ref{F1}a). Recent studies have indicated that such dipole near-field enhancement\cite{tokel2017chip,kerse2016ablation,asgari2024laser,oktem2013nonlinear} underlies the origin of anisotropic patterns observed in laser nanofabrication\cite{li2024super,li2020fib}. When the size approaches the wavelength ($d \approx \lambda$), multipole resonances could emerge\cite{jahani2016all,babicheva2024mie}, transitioning the scattering behavior to the Mie regime. Under such conditions, the light field distribution is strongly modulated and confined by the air-filled void, and displays pronounced wavelength sensitivity\cite{hentschel2023dielectric,hamidi2025quasi}. As shown in Fig.~\ref{F1}b, the degree of light energy confined within the cavity is quantified by integrating the density of electromagnetic energy within the volume of the nanocavity. When the geometry of the laser-ablated initial cavity on the silicon substrate measures approximately 250 nm in width and 100 nm in depth, the cavity exhibits a strong resonant response in the ultraviolet range, leading to a pronounced enhancement of the light near-field. This insight inspired us that, by correctly choosing the processing laser wavelength, Mie resonance itself can dominate the energy deposition and guide the subsequent fabrication process.

We experimentally validated this resonance-driven hypothesis by comparing ablation on the silicon substrate using 800~nm and 343~nm femtosecond lasers. Both lasers initially create similar seed cavities (approximately 250~nm wide, 100~nm deep). However, their subsequent evolution diverges dramatically, as shown in Fig.~\ref{F1}e and g. Under 800~nm laser beam irradiation, the growth of cavity depth was severely suppressed, whereas the lateral size increased dramatically. In sharp contrast, the 343~nm laser beam drives a significant vertical ablation, reaching a depth of $\sim$450~nm with minimal lateral expansion (Fig.~\ref{F1}i and j). This result is highly counterintuitive from an absorption perspective. The 800~nm light has a large optical skin depth in silicon ($\sim$9~$\mu$m), whereas the 343~nm light is absorbed within a mere 9~nm (Fig.~\ref{F1}h). Given this large difference in optical skin depth, conventional models would expect that the 800~nm laser beam should drill much deeper, contrary to the fact that depth growth was suppressed.
The explanation for this paradox lies in the resonant feedback mechanism, confirmed by numerical simulations (Fig.~\ref{F1}d and f). For a 300~nm wide, 150~nm deep cavity, the 343~nm laser beam excites a strongly localized Mie mode within the void. This resonance funnels energy directly to the cavity base, maximizing the energy deposition for material removal. Conversely, the 800~nm laser beam is non-resonant with this structure. It primarily forms a standing wave above the silicon surface due to strong Fresnel reflection, scattering energy away from the cavity base and inhibiting deep drilling (Supplementary Movie~2).

This mechanism reveals a sustaining light confinement phenomenon: (1) the 343 nm laser beam ablates a seed cavity, which supports fundamental Mie resonance in the UV range, confining the field at its base (Fig.~\ref{F1}f and g). This intense near-field enhancement drives selective, vertical ablation (Supplementary Movie~1); (2) as the cavity deepens, its geometry evolves, but it transitions through the excitation of a series of higher-order eigenmodes (Fig.~\ref{F1}c, f and Suppl. Section \blue{2.2}) that also maintain strong light field confinement at the cavity base. This dynamic interplay—where the evolving geometry reshapes the resonant mode, which in turn guides the ablation—ensures the stable, continuous progress of localized laser drilling.

\begin{figure}[H]
\centering
\includegraphics[width=0.87\textwidth]{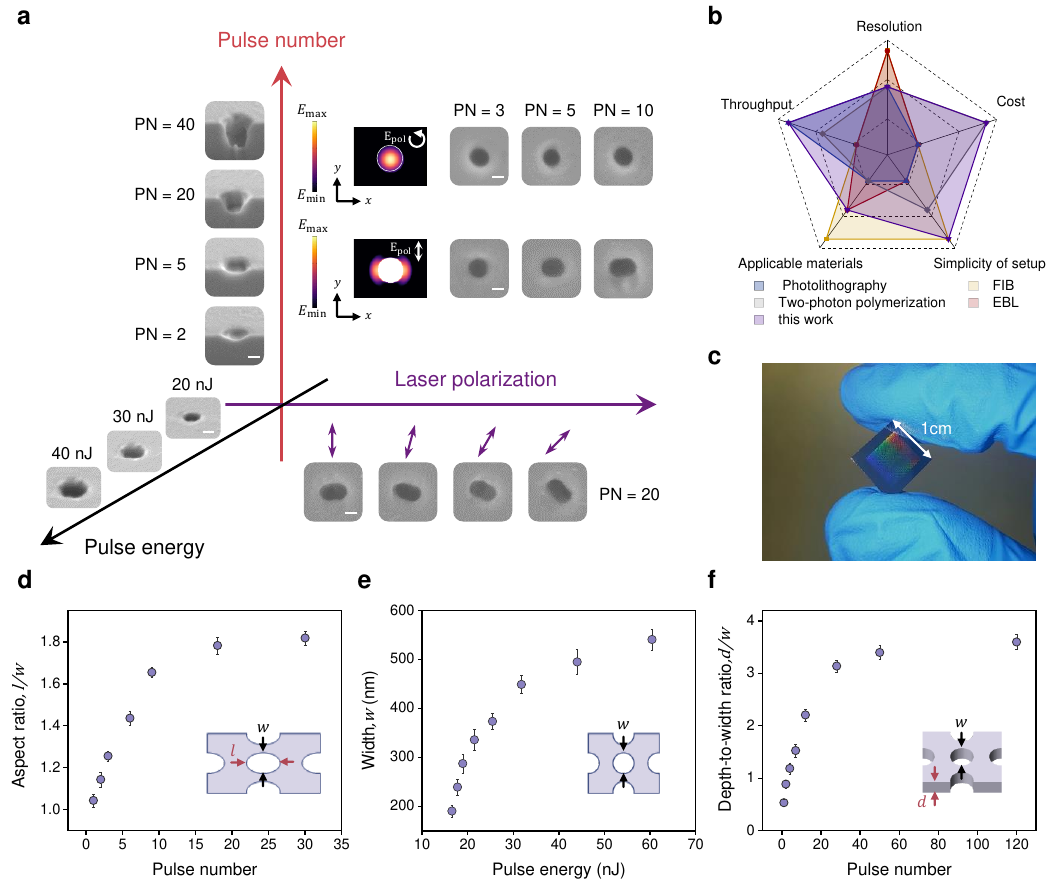}
\vspace{-0cm}
\caption{\textbf{Controllable printing of Mie resonators via Mie-lithography}. \textbf{a} Controlled growth of Mie resonators achieved by varying laser pulse energy, polarization and pulse number. The depth of the resonators increases with the increasing pulse number, while higher pulse energy leads to larger ablation cavity width. The polarization state of the laser beam determines both the symmetry and the orientation of the resulting Mie resonators. Scale bars are all 200 nm. \textbf{b} Performance comparison between Mie lithography and four established nanofabrication techniques in terms of cost, resolution, applicable materials, throughput, and simplicity of setup (detailed parameters can be found in Suppl. Section \blue{1}). The comparison is intended solely to contextualize the performance advantages of Mie-lithography relative to established nanofabrication methods. \textbf{c} Optical image of a large-scale array of Mie resonators fabricated via Mie lithography. \textbf{d}-\textbf{f} show the influence of laser pulse number and pulse energy on the aspect ratio, width and depth-to-width ratio of the Mie resonators.}
\label{F2}
\end{figure}

Mie-lithography provides precise control over the resonators' geometry by tuning laser parameters (Fig.~\ref{F2}a). The pulse number predominantly tunes the cavity depth with a weak influence on its width, whereas the pulse energy primarily determines the width while also influencing the depth evolution, and the polarization state governs the in-plane symmetry of the cavity (Fig.~\ref{F2}d–f). Specifically, circularly polarized light incidence excites circularly symmetric Mie modes, resulting in isotropic ablation cavities. In contrast, under linear polarization, the anisotropic field enhancement induces asymmetric resonators with their major axis aligned perpendicular to the linear polarization. The combined control of pulse number, pulse energy, and polarization orientation provides a multidimensional morphology space with tunable cavity depth, width, and in-plane orientation. Nevertheless, owing to the intrinsic symmetry of the underlying Mie modes, the accessible geometries are currently restricted mainly to circular and elliptical cavities and do not include arbitrary free-form three-dimensional architectures. This fabrication method directly addresses the challenges outlined above (Fig.~\ref{F2}b). It is a single-step process operating in ambient air, capable of achieving high throughput ($\ge 10^6$ pixels/s) with a minimum of just two pulses per resonator and a 10 MHz laser. This allows for scalable production over macro areas (e.g., 1 cm $\times$ 1 cm, Fig.~\ref{F2}c) at low cost, bypassing slow, expensive electron/ion beam etching. Furthermore, the principle relies on refractive index contrast, suggesting potential applicability to other high-index dielectric or semiconductor materials (Suppl. Section \blue{5}), though practical implementation requires re-optimization of the laser parameters.

\subsection{Void-type Mie resonators for ultra-broadband dispersion engineering} 
\begin{figure}[H]
\centering
\includegraphics[width=0.8\textwidth]{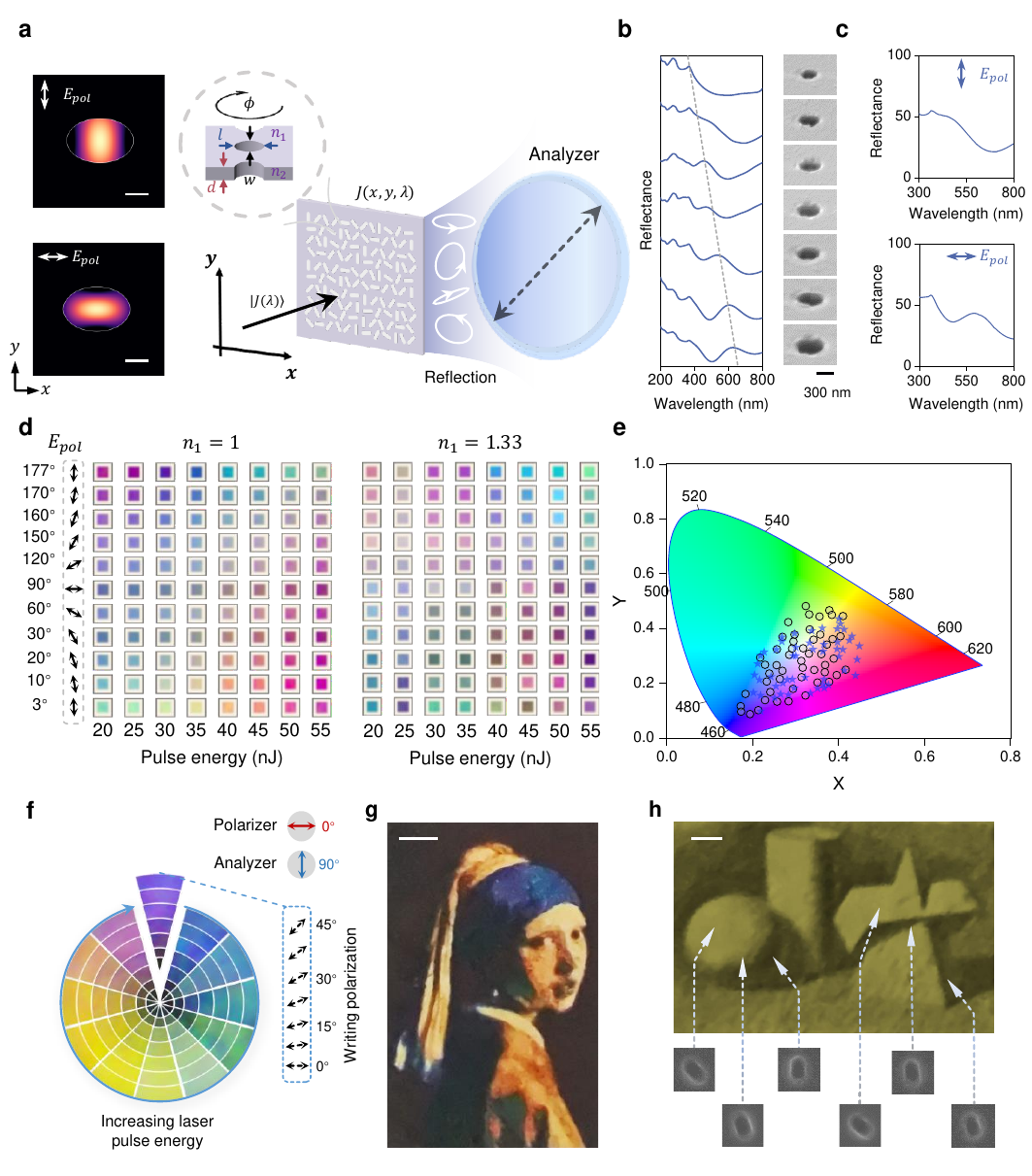}
\vspace{-0.3cm}
\caption{\textbf{Ultra-broadband light field manipulation with Mie resonators}. \textbf{a} Schematic of light field modulation with Mie resonators under reflection mode. Key parameters governing the light-matter interaction include: the length ($l$), width ($w$), orientation ($\phi$) and depth ($d$) of the asymmetric Mie resonators, refractive index contrast between the interior ($n_1$) and exterior ($n_2$) of the resonators. The asymmetric Mie resonators exhibit optical anisotropy to linearly polarized light, with an analyzer extracting distinct polarization components. The repeated resonators represent the experimental sample layout. Scale bars are 200 nm. \textbf{b} Experimentally measured reflectance of Mie resonators on the silicon substrates. The measured geometric parameters (width, depth; nm) are as follows: (290, 200), (335, 280), (380, 330), (415, 400), (440, 435), (480, 505), (530, 570). \textbf{c} Experimental characterization reveals polarization-dependent resonant spectral shifts. \textbf{d} shows the Mie lithography-fabricated color palettes experimentally recorded in air ($n_1=1$) and water ($n_1=1.33$), which exhibit color tuning with increasing polarization angle. Asymmetric Mie resonators oriented at 45° were fabricated with a fixed pulse number (PN = 20) and pulse energies varied from 20 nJ to 55 nJ. The pulse energies are measured as the pre-objective energy. Black squares were used to guide the eye. During optical characterization, the analyzer angle was maintained at 90°. \textbf{e} CIE chromaticity diagram comparing colors of silicon-based Mie resonators in water (stars) and in air (circles). \textbf{f} Chromatic wheel with continuously adjustable brightness with orthogonal polarizer–analyzer combinations. \textbf{g} The microscope image of the oil painting \textit{Girl with a Pearl Earring}. \textbf{h} The microscope image of a still-life sketch. Mie resonators with different orientations exhibit distinct shading states. The length and width of each resonator are 650~nm and 350~nm, respectively. Scale bars in \textbf{g} and \textbf{h} are 50~$\mu$m.}\label{F3}
\end{figure}

Having elucidated the fabrication mechanism of Mie-lithography, we now demonstrate how these resulting void-type Mie resonators enable ultra-broadband light modulation. Mie voids have previously been explored across the terahertz region for high-$Q$ sensing applications\cite{kumar2025extremely}. Here, we exploit their size-dependent optical responses to realize continuous spectral modulation from the deep-ultraviolet to the near-infrared. Through judicious parameter selection (pulse energy and pulse number), we successfully fabricated Mie resonators covering the deep-ultraviolet to the near-infrared range. As shown in Fig.~\ref{F3}a and b, increasing the resonator size induces a redshift of the resonance peaks, along with the emergence of multiple peaks corresponding to higher-order modes (see Suppl. Section \blue{2.2} for details). These observed higher-order modes also corroborate our fabrication model: the fact that these resonators strongly support multiple UV resonant modes validates the “resonance-in-air” feedback mechanism (driven by the 343 nm laser beam) proposed in the previous section.

The asymmetric resonators exhibit polarization-dependent electric field distributions (Fig.~\ref{F3}a) and produce the phase difference under orthogonal polarization excitation\cite{arslan2026polarization}, which can be regarded as nanoscale wave-retarders (Fig.~\ref{F3}c and Suppl. Section \blue{7}). Leveraging this mechanism, we demonstrate dynamic color switching controlled by the polarization of incident white light. For instance, a Mie resonator (approximately 400~nm long, 300~nm wide and 200~nm deep) fabricated with a single-pulse energy of 20 nJ and 20 pulses exhibits a color shift from bluish green to purplish red as the incident polarization angle rotates from 3° to 177° (Fig.~\ref{F3}d and see Suppl. Section \blue{7.2} for details). Furthermore, because the resonant mode is confined in air, it is exceptionally sensitive to the cavity's refractive index ($n_1$). When the resonator of the above dimensions (size: 400~nm × 300~nm × 200~nm) is immersed in water, its color under 177° polarized illumination shifts from purplish red to red. Other resonators also exhibit notable optical response variations (Fig.~\ref{F3}d, e). This pronounced refractive-index-dependent behavior establishes them as highly promising candidates for nanoscale optical sensing platforms to detect liquids\cite{arslan2025attoliter}, high-pressure refractive-index variations\cite{niinomi2025mie}, and nanoplastics\cite{ludescher2025optical}. 

The color manipulation capability of asymmetric Mie resonators readily extends across the hue, saturation and value (HSV) color space. Color brightness can be easily modulated by rotating the resonator's orientation under orthogonally polarized incident and reflected light (see Suppl. Section \blue{7.3} for detailed modulation principles). Although previous studies have reported related phenomena based on plasmonic resonances\cite{song2023versatile}, our work exploits Mie resonance to simultaneously tune hue, saturation and brightness value within individual resonators, yielding a resolution (see Suppl. Section \blue{6} for details) of up to 63,500 DPI—an order of magnitude higher than prior comparable work\cite{song2023versatile,zhang2023micro}. Fig.~\ref{F3}f presents an optical image of a chromatic wheel fabricated via Mie lithography. Each radial sector of the wheel consists of Mie resonators with identical sizes but varying orientations (rotated from 0° to 45°), resulting in smooth brightness gradients within the same color. By encoding color and brightness information into Mie resonators with tailored sizes and orientations, printed artworks (Fig.~\ref{F3}g and h) achieve stereoscopic effects under white-light LED illumination.

\subsection{Direct laser printing of DUV-NIR nano dispersion devices} 
\begin{figure}[H]
\centering
\includegraphics[width=0.85\textwidth]{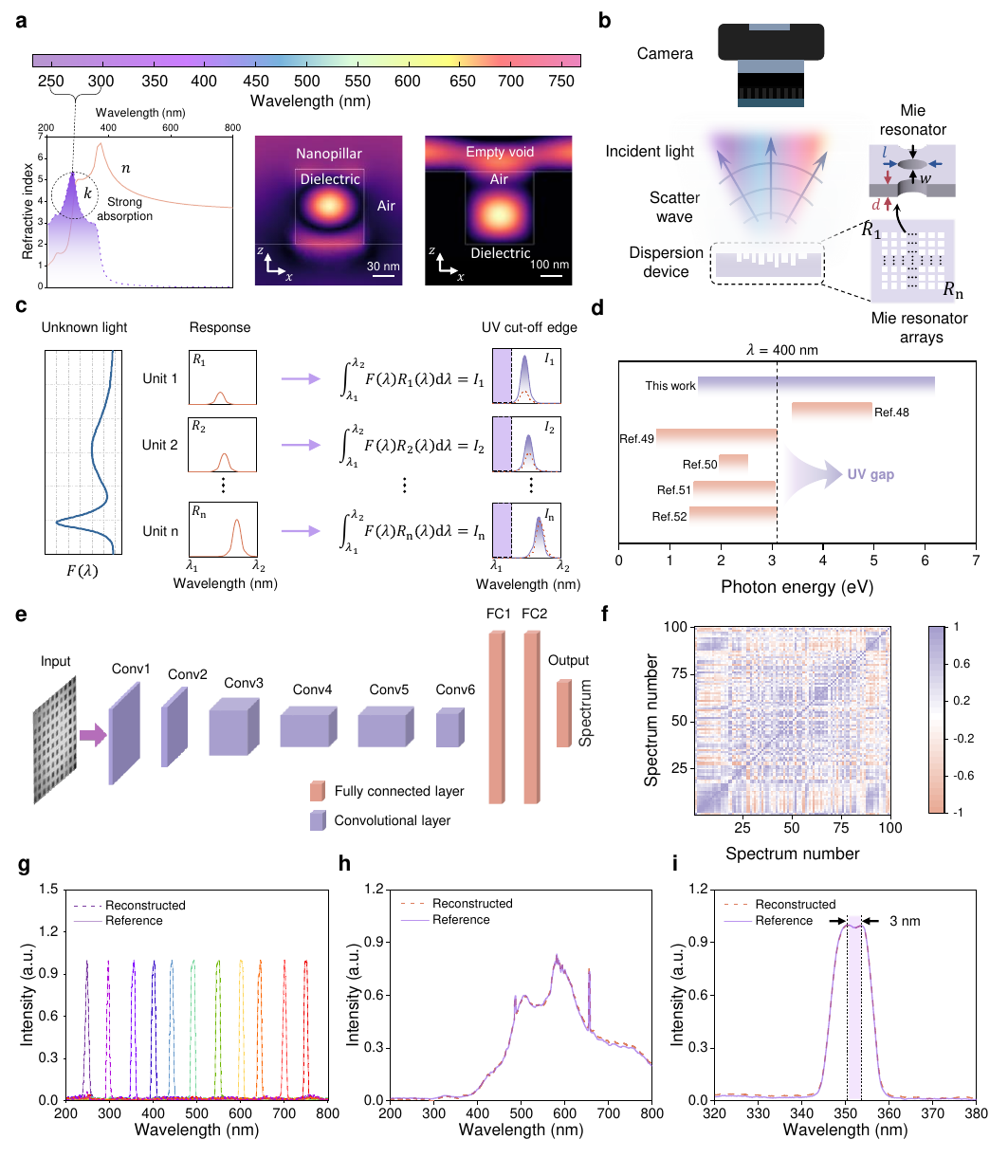}
\vspace{-0.4cm}
\caption{\textbf{DUV-NIR nano dispersion devices based on resonance in air}. \textbf{a} Current miniaturized spectrometers utilize dispersion generated by resonance in high-refractive-index nanopillars or multilayer thin films, which is constrained by material absorption, particularly in the ultraviolet region. In contrast, our ``resonance-in-air'' concept effectively alleviates the strong absorption inherent to high-index materials from DUV to NIR. \textbf{b} illustrates the principle of encoding spectral information based on resonance. Each unit in the resonator arrays is engineered to support diverse spectral responses R($\lambda$). When the incident light with an unknown spectrum F($\lambda$) is modulated by nano resonators, the encoded optical signal ($I$) is captured in a single camera exposure. However, the weak spectral response in the UV range results in a severely attenuated signal that is often dominated by noise (\textbf{c}). \textbf{d} A comparison of the operating wavelength ranges between our device and reported miniaturized spectrometers\cite{yu2025miniaturized,bian2024broadband,yang2019single,yoon2022miniaturized,fan2024dispersion}. \textbf{e} Schematic of the neural network for spectral reconstruction. Input: the reflected signal from our fabricated nano dispersion device, captured by a UV camera. Output: reconstructed spectra covering the 200–800~nm wavelength range. \textbf{f} Correlation coefficient matrix computed from the reflectance of 100 resonator arrays in our nano dispersion device. \textbf{g} Narrowband spectral reconstruction results compared with measurements from a commercial spectrometer. \textbf{h} Complex spectral reconstruction. \textbf{i} Two spectral peaks separated by approximately 3 nm can be resolved in the reconstructed spectrum.}\label{F4}
\end{figure}

Dispersion engineering is crucial not only for color reproduction but also for enabling customized dispersion in miniaturized spectrometers\cite{zhang20253d,wang2019single,he2024meta,tang2024metasurface,bao2015colloidal,cai2024compact}. The miniaturization of such devices is empowered by the localized dispersion from optical subwavelength structures. As discussed in the introductory part, existing architectures and materials have yet to simultaneously meet the demands of low optical loss and strong dispersion across an ultra-broadband spectral range, especially in the DUV-NIR range (Fig.~\ref{F4}a). As shown in Fig.~\ref{F4}b and c, the incident spectrum$\ F(\lambda)$ is modulated by nanostructures with distinct dispersive responses (e.g., transmittance and reflectance; specifically, the reflectance is utilized in our case). Then, the resulting intensity captured by a camera can be expressed as: $I_i=\int R_i\left(\lambda\right)F(\lambda)\mathrm{d}\lambda$, where $R_i\left(\lambda\right)$ represents the response of the $i_{th}$ sub-array. Owing to material absorption, conventional miniaturized spectrometers suffer from weak UV signals that are often overwhelmed by noise, limiting most miniaturized systems to the visible and near-infrared range (Fig.~\ref{F4}d). Despite being the first demonstration of a UV micro-spectrometer based on AlGaN n–p and GaN p–n diodes\cite{yu2025miniaturized}, the devices developed by Yu et al. have an operational bandwidth limited to 250–365 nm. This limitation hinders their ability to bridge the ultraviolet-to-visible spectral gap (Fig.~\ref{F4}d).
To address this gap, we utilize resonances confined within low-loss air voids to alleviate the strong intrinsic absorption of high-index materials. Specifically, we fabricated a device comprising 10 $\times\ $10 arrays of Mie resonators (approximately 100 $\mu$m $\times\ $100 $\mu$m in total area) via Mie lithography. Each array was designed to exhibit a distinct reflectance (typical reflectance is provided in Suppl. Section \blue{9.1}), while the correlation coefficients are shown in Fig.~\ref{F4}f. These scattered signals from the Mie resonator arrays can be captured simultaneously in a single-shot exposure. Combined with an optimized convolutional neural network (Fig.~\ref{F4}e and details in Suppl. Section \blue{9.2}), this approach enables ultra-broadband spectral reconstruction from DUV to NIR within milliseconds. To validate our concept, a set of single-peak spectra was reconstructed across the 200–800 nm range. The spectra align well with reference measurements from a commercial spectrometer (Fig.~\ref{F4}g) with a reconstruction error of $\varepsilon_\lambda=$ 1.57\% ($\varepsilon_\lambda=||\lambda_0-\lambda_R||_1/||\lambda_0||_1$), in which $\lambda_0$ and $\lambda_R$ are the ground truth and the reconstructed spectrum, respectively. The system also accurately reconstructs complex spectra (Fig.~\ref{F4}h), achieving a comparable reconstruction error $\varepsilon_\lambda$ of 1.43\% and a spectral resolution of 3~nm (defined as the minimum wavelength separation at which two adjacent spectral peaks remain distinguishable as two local maxima in the reconstructed spectrum). 

\subsection{Conclusion and outlook} In conclusion, this work addresses the single-material limitations imposed by the Kramers–Kronig relations in ultra-broadband dispersion devices and develops a nanofabrication technique for dispersion devices that possesses large-scale scalability potential, using a unified strategy that shifts the light-matter interaction from solid structures to air-filled voids. Specifically, through a feedback mechanism between a UV laser beam and cavities, we achieved the fast printing of three-dimensional void-type resonators exhibiting optical dispersion from DUV to NIR. As a proof of concept, we realized color printing with exceptional resolution up to 63,500 DPI and a miniaturized spectral chip operating from 200 nm to 800 nm---a bandwidth unprecedented in compact devices. Although the current reconstruction is limited by the camera, the bandwidth of the resonators is not constrained and can be extended into the infrared by increasing the laser energy. This single-step printing strategy demonstrates remarkable universality, enabling the fabrication of diverse dispersion devices, such as data storage and anti-counterfeiting (see Suppl. Section \blue{8}). Our method thus provides a high-throughput, low-cost fabrication route for scaling future ultra-broadband dispersion devices.


\begin{figure}[H]
\centering
\includegraphics[width=0.87\textwidth]{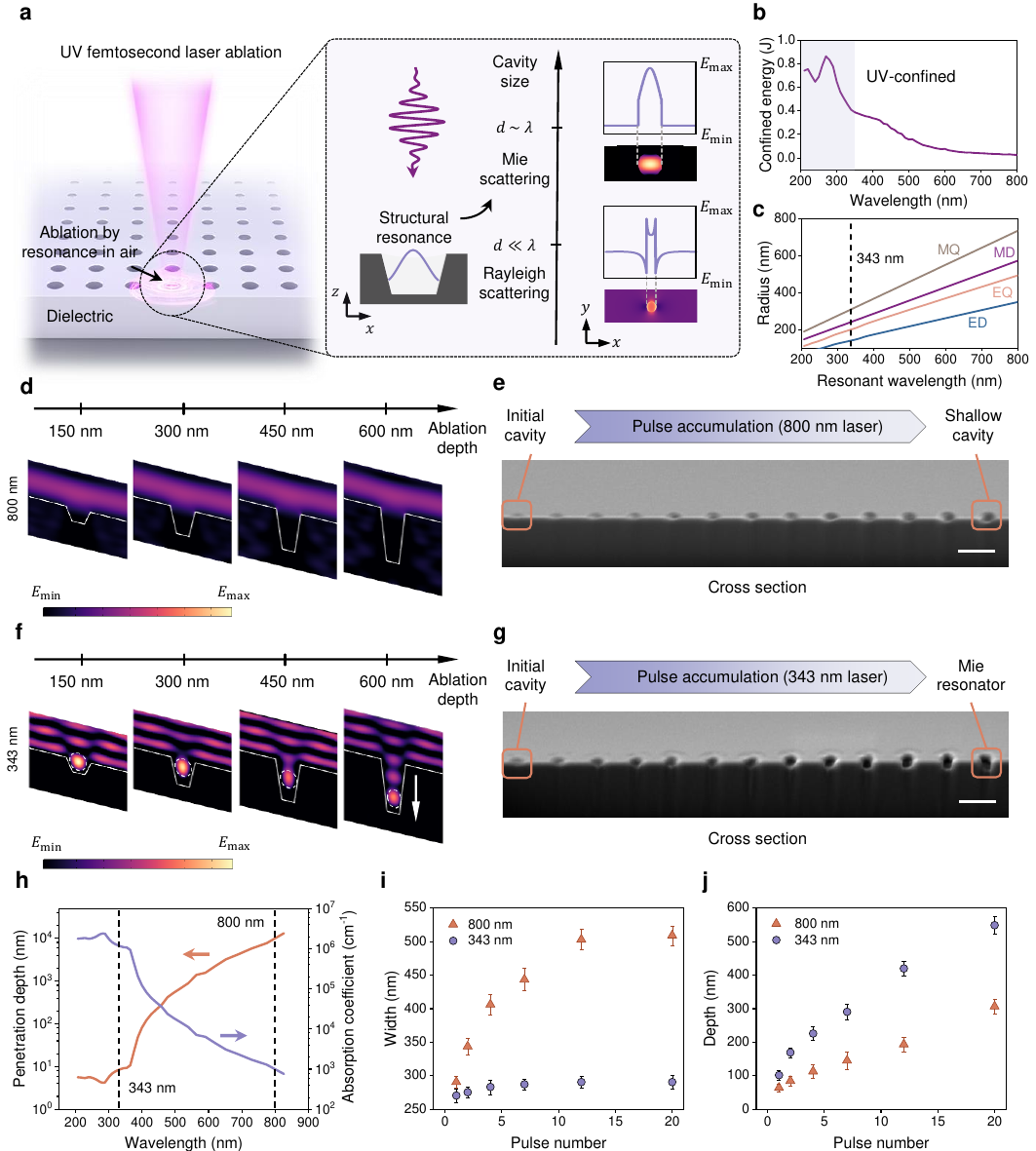}
\vspace{-0cm}
\caption{\textbf{Mechanism of Mie-lithography}. \textbf{a} The schematic illustrates the localized resonant modes excited by cavities of different sizes, which drive a feedback loop between the incident laser beam and the laser-ablated nanocavities. The cavity profiles in the right panel are simplified schematic representations used to illustrate the resonant mechanism and field distributions. \textbf{b} Simulation results demonstrate that the initial cavity (250~nm width and 100~nm depth) formed by laser ablation provides strong light confinement for ultraviolet lasers. \textbf{c} Multipole analysis of Mie resonators. As the size of Mie resonators increases, higher-order multipole modes emerge within the UV spectral range, thereby sustaining the feedback loop. $\mathrm{ED}$: electric dipole, $\mathrm{EQ}$: electric quadrupole, $\mathrm{MD}$: magnetic dipole, $\mathrm{MQ}$: magnetic quadrupole. \textbf{d}, \textbf{f} Electric field intensity profiles of cavities (fixed cavity width of 300~nm, depths varying from 150 to 600~nm) under 800~nm (\textbf{d}) and 343~nm (\textbf{f}) laser beam illumination. The simulations reveal wavelength-dependent light-localization trends. \textbf{e}, \textbf{g} Experimental verification (Silicon) that Mie resonance excited by the 343~nm laser beam drives self-assembly growth of Mie resonators. At the non-resonant wavelength of 800~nm, axial growth is suppressed due to strong scattering despite silicon's larger skin depth (\textbf{h}) at 800~nm. \textbf{i}, \textbf{j} The influence of laser wavelength on cavity evolution with increasing pulse number. The 343~nm laser beam enables direct fabrication of nanostructures with a high depth-to-width ratio, which is attributed to the feedback mechanism that tightly confines light at the cavity bottom. Scale bars in \textbf{e} and \textbf{g} are $1~\mu\mathrm{m}$.}
\label{F1}
\end{figure}
\newpage

\clearpage

\begin{methods}
\subsection{Laser fabrication system} In the experiments, we utilized two femtosecond laser systems: the Light Conversion Carbide (pulse duration: 220~fs, center wavelength: 1030~nm, maximum repetition rate: 1 MHz) and the Spectra-Physics Spitfire (pulse duration: 150~fs, center wavelength: 800 nm, repetition rate: 1 kHz). The Carbide laser was converted to 343 nm via the harmonic generator (HIRO, Light Conversion). The laser pulses were tightly focused onto the sample surface through a high-numerical-aperture objective lens (NA = 0.95), ensuring a submicrometer scale beam waist radius of the focused light spot.
The laser beam was scanned across a 100 $\times$ 100 $\mu\mathrm{m}^{2}$ processing area using a Galvano-mirror XY scanner. The energy for laser processing was modulated using a combination of a zero-order half-wave plate and a polarizing beam splitter to achieve 0-100\% continuous energy attenuation. An additional zero-order half-wave plate driven by a high-precision stepper motor rotary stage (MRS312, Beijing Optical Century Instrument) enabled programmable polarization rotation from 0° to 180°.

\subsection{Sample characterization} To characterize the three-dimensional surface topography of the Mie resonators, we employed a focused dual-beam field emission scanning electron microscope (FEI-Scios 2 HiVac) to perform cross-sectioning of the Mie resonators (Fig.~\ref{F1}e and g) while simultaneously acquiring scanning electron microscopy images. Prior to imaging, the samples were ultrasonically cleaned in a 20\% hydrofluoric acid (HF) aqueous solution. The artwork was characterized using an optical microscope (BX53, Olympus). Samples were fixed on an XYZ stage and illuminated by a broadband LED source (400–700 nm) through a low-magnification objective (MPLFLN10XBD, NA = 0.25, Olympus). 
The reflectance of Mie resonators was characterized using a home-built micro spectroscopy setup. Broadband light from a deuterium-halogen (190$-$400~nm $\&$ 360$-$2500~nm) combined source (LBDH2000,
LBTEK) was focused onto the sample through a DUV objective lens (M Plan UV 20x, Mitutoyo). The reflected signal was then collected by the same objective and coupled into a fiber-optic spectrometer (AvaSpec-ULS4096CL-EVO, Avantes) for spectral analysis. Simultaneous imaging was performed using a UV camera (Alvium 1800 U-812 UV). A detailed schematic of the optical setup is provided in the Supplementary Information.

\subsection{Numerical simulations} Using COMSOL Multiphysics 6.0, we simulated the electric field distributions under 343 nm and 800 nm laser beam irradiation for cavities of varying sizes (depth: 150–600 nm), as shown in Fig.~\ref{F1}d and f. For numerical
simulations of the eigenfrequencies of cavities with different sizes (Fig.~\ref{F1}c), we performed calculations using our custom Python code based on Mie theory. Relevant discussions are provided in Suppl. Section \blue{2}.
\end{methods}

\begin{addendum}
 \item 
The work was funded by the Scientific and Technological Innovation Project of Changbaishan Laboratory, Jilin Province (Grant No. CBS2026007-02); National Natural Science Foundation of China (62175086; 62505159; U2541221). Z.-Z.L. is grateful for the support by the Youth Talent Promotion Project of the China Association
for Science and Technology, Shuimu Tsinghua Scholar Program, and China Postdoctoral Science Foundation, 2025T180228 and 2024M761640.

 \item[Author contributions]
W.G., Z.-Z.L., L.W., and H.-B.S. conceived the concept. W.G., C.Y., and Z.W. performed the experiments. Z.-Z.L., W.G., Y.W., L.W., C.-Q.J. and H.-B.S. performed the theoretical analysis and improved the results. W.G., C.Y., H.-Y.F., Y.-H.L. and Z.W. contributed to the characterization. Z.-Z.L., Q.-D.C., L.W., and H.-B.S. supervised the whole project. W.G., Z.-Z.L., and H.-B.S. wrote the initial draft and all authors contributed to the final paper.

 \item[Competing financial interests]
 The authors declare that they have no conflict of interest.
  \item[Data availability] All data needed to evaluate the conclusions in the paper are available in the main text or the supplementary materials.
 \item[Additional information]
 Supplementary Information is available for this paper. Correspondence and requests for materials should be addressed to Z.-Z. Li, L. Wang, Q.-D. Chen, or H.-B. Sun.
\end{addendum}

\nolinenumbers
\newpage
\noindent\textbf{References}

\bibliographystyle{naturemag}
\bibliography{sample}

\end{document}